Title page

**Title:** Independent Action Models and Prediction of Combination Treatment Effects for Response Rate, Duration of Response and Tumor Size Change in Oncology Drug Development

**Authors and affiliations**: Linda Z. Sun[a,*], Cai (Iris) Wu[a], Xiaoyun (Nicole) Li[a], Cong Chen[a], and Emmett V. Schmidt[b]

[a] Biostatistics and Research Decision Sciences, Merck & Co., Inc., Kenilworth, NJ 07033, USA

[b] Oncology Early Development, Merck & Co., Inc., Kenilworth, NJ 07033, USA

[*] Corresponding author: linda_sun@merck.com.

**Acknowledgement**

The authors sincerely thank Dr. Eric H. Rubin from Merck & Co., Inc., Kenilworth, NJ USA for critical review and insightful comments and suggestions.

# Abstract

An unprecedented number of new cancer targets are in development, and most are being developed in combination therapies. Early oncology development is strategically challenged in choosing the best combinations to move forward to late stage development. The most common early endpoints to be assessed in such decision-making include objective response rate, duration of response and tumor size change. In this paper, using independent-drug-action and Bliss-drug-independence concepts as a foundation, we introduce simple models to predict combination therapy efficacy for duration of response and tumor size change. These models complement previous publications using the independent action models (Palmer 2017, Schmidt 2020) to predict progression-free survival and objective response rate and serve as new predictive models to understand drug combinations for early endpoints. The models can be applied to predict the combination treatment effect for early endpoints given monotherapy data, or to estimate the possible effect of one monotherapy in the combination if data are available from the combination therapy and the other monotherapy. Such quantitative work facilitates efficient oncology drug development.

2
# Abstract

An unprecedented number of new cancer targets are in development, and most are being developed in combination therapies. Early oncology development is strategically challenged in choosing the best combinations to move forward to late stage development. The most common early endpoints to be assessed in such decision-making include objective response rate, duration of response and tumor size change. In this paper, using independent-drug-action and Bliss-drug-independence concepts as a foundation, we introduce simple models to predict combination therapy efficacy for duration of response and tumor size change. These models complement previous publications using the independent action models (Palmer 2017, Schmidt 2020) to predict progression-free survival and objective response rate and serve as new predictive models to understand drug combinations for early endpoints. The models can be applied to predict the combination treatment effect for early endpoints given monotherapy data, or to estimate the possible effect of one monotherapy in the combination if data are available from the combination therapy and the other monotherapy. Such quantitative work facilitates efficient oncology drug development.

# 1. Introduction

With advances in our knowledge of cancer genetics and relevant target pathways, there are unprecedented numbers of new targets in development for cancer therapy. At the same time, single agent targeting has rarely shown adequate efficacy to improve upon existing stand of care (SoC) therapies, especially in broader populations. Therefore, almost all investigational agents are developed in combination therapy. Particularly since PD-1 checkpoint inhibitors have become the SoC of many tumor types, thousands of clinical trials are ongoing to test combination therapies using PD-1 checkpoint inhibitors as the backbone. [17, 18, 22]

Response rate together with duration of response (DoR), and tumor size change are the most used early endpoints in early oncology development to decide which combinations to move forward to late stage development. The durable responses from immunotherapies are considered a breakthrough as they increase a patient's chance to achieve long-term survival benefit [15]. Many recent accelerated approvals of the PD-1 checkpoint inhibitors from FDA are based on response rate and durability of response. When the PD-1 checkpoint inhibitors are combined with other drugs (e.g. chemotherapy, targeted therapy, or another immune checkpoint inhibitor), will the duration of response be further extended? What would be the expected response rate of the combination therapy? Will the combination induce deeper response, which correlates with longer survival [8, 16]?

To answer these questions, in Section 2, we first introduce the concepts of independent drug action (IDA) and Bliss drug independence (Bliss). We then apply these principles to predict response rates for combination therapies. In Section 3, we study the nature of responders to combination therapies, dissect the responders into 3 categories, and apply the independent action model to predict DoR for combination treatments. Then in Section 4, using the same categorization of responders, we propose a model to predict combination treatment effect for tumor size change modified from the independent drug action and Bliss, which may improve upon the original independent drug action model. In each section, the predictions are compared with observed trial data to demonstrate that the proposed models provide reasonable predictions and are easy to implement. Statistical inferences, some practical recommendations in prediction, and potential application in decision making are provided.

## 2. Independent Drug Action, Bliss Model, and Response Rates for Combinations

The independent drug action model assumes that, in a combination therapy, each patient benefits from the drug to which his or her tumor is most sensitive [16]. From this definition, one can see that the model does not require that the drugs in the combination hit completely separate targets. The model allows correlation among the drugs. With the assumption of independent drug action, administering multiple drugs to patients is a bet-hedging strategy which increases each patient's chance of experiencing a clinically meaningful anti-tumor response to any single drug in the combination. The anti-tumor response can be measured as response (e.g. complete response or partial response per RECIST in solid tumors), duration of response, tumor size reduction or progression-free survival (PFS), etc. Palmer 2017 [17], Palmer 2020 [18] and Schmidt 2020 [22] showed that, due to high inter-patient variability in humans, independent drug action model explains the observed treatment effect for majority of combination therapies in terms of PFS and objective response rate (ORR).

When applying the independent drug action model to predict the ORR of a combination from the individual drug's ORR, let $X_1$ and $X_2$ be a patient's response to Drug 1 and Drug 2, respectively. $X_i = 1$ ($i = 1, 2$) means the patient achieves complete response (CR) or partial response (PR) according to RECIST 1.1., and $X_i = 0$ means no response is achieved. Let $r_i = \Pr(X_i = 1)$ ($i = 1, 2$) be the ORR to the individual drugs. According to the independent drug action model, a patient's response to the combination of Drug 1 and Drug 2 is $X = \max(X_1, X_2)$. The ORR $r$ to the combination therapy is predicted by (1) (See Appendix 1 for the derivation)

$$r = \Pr(X = 1) = r_1 + r_2 - r_1 r_2 - \varphi' \sqrt{r_1(1 - r_1)r_2(1 - r_2)} \tag{1}$$

In equation (1), $\varphi'$ is the correlation between $X_1$ and $X_2$. If $X_1$ and $X_2$ are independent (i.e. $\varphi' = 0$), $r = r_1 + r_2 - r_1 r_2$.

In parallel, Schmidt 2020 [22] used Bliss Drug Independence Model to predict the ORR of a combination therapy. The Bliss Model originates at the cell level. If Drug 1 kills $m_1$ proportion of tumor cells, and Drug 2 kills $m_2$ proportion of tumor cells, then based on the probabilistic independence $m_1 + m_2 - m_1 m_2$ proportion of tumor cells will be killed by the combination if the two drugs work independently [2]. When the Bliss model is applied to the patient level, if Drug 1 induces response in $r_1$ proportion of patients, and Drug 2 $r_2$, then the combination is predicted to have response rate of $r_1 + r_2 - r_1 r_2$. Therefore, the Bliss Drug Independence Model is a special case (i.e. $\varphi' = 0$) of the independent drug action model in predicting response rate of a combination therapy. The prediction from the Bliss Drug Independence model can be a benchmark to evaluate possible synergist effects between Drug 1 and Drug 2. If the observed response rate of the combination is higher than the prediction, it may indicate that synergy exists. Schmidt 2020 [22] showed that the predicted response rates for the combinations using Bliss model are close to the observed response rate for the PD-1 checkpoint inhibitor combinations in 319 results from 98 studies.

In the following sections we use two Phase III studies, Keynote 062 and Checkmate 067, as examples for the prediction models, and here we illustrate the response rate prediction for these two studies in this section. Keynote 062 and Checkmate 067 are ideal studies to apply the prediction models because the combination arm and the two monotherapy arms are in the same trial with randomization. Keynote 062 [23] is a randomized, controlled, partially blinded Phase 3 study of patients with untreated, locally advanced/unresectable or metastatic G/GEJ cancer with PD-L1 CPS of 1 or greater. Patients were randomized 1:1:1 to pembrolizumab, pembrolizumab plus chemotherapy, or chemotherapy. In Keynote 062, the response rate is 14.8% in the monotherapy pembrolizumab arm and 37.2% in the chemotherapy arm. From equation (1), the predicted response rate of the combination is 46.5% assuming $\varphi' = 0$, which is close to the observed response rate in the combination arm 48.6%. Checkmate 067 [12] is a Phase 3 trial of patients with previously untreated stage III/IV melanoma. Patients were randomly assigned in a 1:1:1 ratio to ipilimumab, nivolumab or ipilimumab plus nivolumab. In Checkmate 067, the response rate is 19.0% in the monotherapy ipilimumab arm and 43.7% in the monotherapy nivolumab arm. The predicted response rate for the combination is 54.4% assuming $\varphi' = 0$, which is close to the observed response rate in the combination arm 57.6%.

## 3. Duration of Response

In addition to response rate, it is also recognized that the duration of response (DoR) is important in contributing to the survival benefit of a cancer therapy. The responses induced by PD-1 checkpoint inhibitors are generally more durable than those of chemotherapies or targeted therapies. This sets high expectations for the DoR of combinations of PD-1 checkpoint inhibitors with other drugs. Intuitively, since the responses are induced by a combination containing a PD-1 checkpoint inhibitor, they must be at least as durable as from PD-1 checkpoint inhibitor alone while the response rate can be boosted by adding another drug. However, it is observed that the median DoR of PD1/PD-L1 in combination with chemotherapy is not as long as the monotherapy PD1/PD-L1, although longer than chemotherapy alone ([Table 1](#) in Appendix 4 shows a few examples).

Why is that? The independent action concept gives a key rationale. In the independent drug action model, the responses induced by a combination of Drug 1 and Drug 2 can be categorized into 3 types:

(1) A patient who would respond to both monotherapy Drug 1 and monotherapy Drug 2.

(2) A patient who would respond to Drug 1 but not to Drug 2.

(3) A patient who would respond to Drug 2, but not to Drug 1.

If Drug 1 is chemotherapy and Drug 2 is a PD1/PDL1 inhibitor, responses to the combination in category (1) and (3) are as durable as monotherapy immunotherapy, but the responses in the combination in category (2) may not be. Therefore, the DoR of the responses induced by the combination is a weighted average of immunotherapy DoR and chemotherapy DoR, which can explain what we observe from Table 1.

To translate the independent action model into mathematical form, let $T_1$ and $T_2$ be the DoR to monotherapy Drug 1 and monotherapy Drug 2, respectively; $T$ be the DoR to the combination of Drug 1 and Drug 2. Let $S_1(t)$, $S_2(t)$ and $S(t)$ be the survival function (i.e. probability of DoR is longer than time $t$) for $T_1$, $T_2$ and $T$, respectively. Figure 1 depicts the model to predict the DoR of a combination therapy.

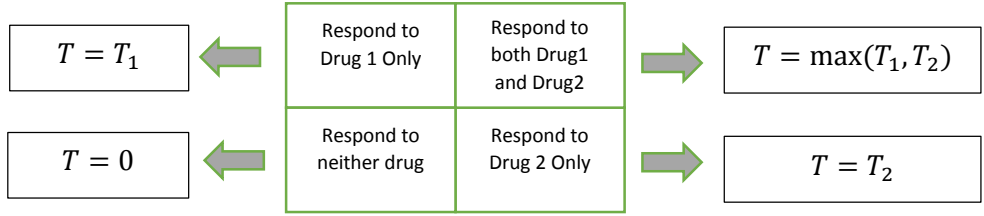

**Figure 1**   Proposed model to predict the duration of response for combination treatment

Let $R$ be the indicator for types of response induced by the combination therapy.

- For patients who would respond to both Drug 1 and Drug 2, $R = Both$. Proportion of such patients among the responders to the combination is $\Pr(R = Both) = r_{12} = \frac{r_1 r_2 + \varphi' \sqrt{r_1(1-r_1)r_2(1-r_2)}}{r}$, where $r$ is from equation (1) in Section 2. For these responders, their DoR is $\max(T_1, T_2)$ under IDA model. The response duration survival function of $\max(T_1, T_2)$ is

$$\Pr(max(T_1, T_2) > t) = \Pr(T_1 > t \ or \ T_2 > t)$$

$$= S_1(t) + S_2(t) - S_1(t)S_2(t) - \varphi''(t)\sqrt{S_1(t)\bigl(1 - S_1(t)\bigr)S_2(t)\bigl(1 - S_2(t)\bigr)}$$

Where $\varphi''(t)$ is the correlation between indicator $I_{T_1>t}$ and $I_{T_2>t}$.

- For patients who would respond only to Drug 1 but not to Drug 2, $R = Drug\ 1$. Proportion of such patients among the responders to the combination is $\Pr(R = Drug\ 1) = r_{10} = \frac{(1-r_2)r_1 - \varphi\prime\sqrt{r_1(1-r_1)r_2(1-r_2)}}{r}$. For these responders, their DoR is $T_1$.

- For patients who response only to Drug 2 but not to Drug 1, $R = Drug\ 2$. Proportion of such patients among the responders to the combination is $\Pr(R = Drug\ 2) = r_{02} = \frac{(1-r_1)r_2 - \varphi\prime\sqrt{r_1(1-r_1)r_2(1-r_2)}}{r}$. For these responders, their DoR is $T_2$.

Therefore, $S(t) = \sum \Pr(T > t|R)\Pr(R)$

$$= \frac{r_{12}\left\{S_1(t) + S_2(t) - S_1(t)S_2(t) - \varphi''(t)\sqrt{S_1(t)(1-S_1(t))S_2(t)(1-S_2(t))}\right\} + r_{10}S_1(t) + r_{02}S_2(t)}{r} \quad (2)$$

Interestingly, if we define $Y_i = X_i T_i$ ($i = 1, 2$) as the DoR for a patient regardless of this patient is a responder or not, i.e. the DoR $= 0$ for non-responders, $\Pr(Y_i > t) = \Pr(X_i T_i > t) = \Pr(T_i > t | X_i = 1)\Pr(X_i = 1) = r_i S_i(t)$. Let $Y = XT$ be the DoR for a patient to the combination therapy, where $X$ is a patient's response to combination therapy and $T$ is the DoR for the responders to the combination therapy. Under independent drug action model $Y = \max(Y_1, Y_2)$. The survival function of DoR for responders to the combination therapy is

$$S(t) = \Pr(T > t) = \frac{\Pr(Y > t)}{\Pr(X = 1)}$$

$$= \frac{r_1 S_1(t) + r_2 S_2(t) - r_1 r_2 S_1(t) S_2(t) - \varphi(t)\sqrt{r_1 S_1(t)(1 - r_1 S_1(t))r_2 S_2(t)(1 - r_2 S_2(t))}}{r} \quad (3)$$

where $\varphi(t)$ is the correlation between indicator $I_{X_1 T_1 > t}$ and $I_{X_2 T_2 > t}$. After incorporating the relationship among $\varphi(t)$, $\varphi'$, and $\varphi''(t)$ as shown in Appendix 2 into the equation, it can be shown that equation (3) and equation (2) are the same. While the derivation of $S(t)$ from $Y = XT$ is quick, easy, and consistent with the derivation to predict combination therapy progression-free survival (PFS) in Chen 2020 [4], the derivation from the nature of the responses to the combination (i.e. dissecting into 3 types) provides more insights of the plausible underlying mechanism of combination effect.

The DoR curve $S(t)$ for the combination therapy can be predicted by (3) from individual drug's DoR curves $S_1(t)$ and $S_2(t)$, and the variance of $\hat{S}$ is shown in Appendix 3. To illustrate the prediction, we apply the model to Keynote 062. Let Drug 1 be the chemotherapy and Drug 2 be pembrolizumab monotherapy. The ORR for Drug 1 and Drug 2 ($r_1$ and $r_2$) and the DOR curves $S_1(t)$ (green line) and $S_2(t)$ (pink line) are reported in Tabernero 2019 [24]. We take a series of $t$ values, e.g. $t$ = 0, 0.5, 1, …, 30 months, and then for each $t$, we predict $S(t)$ from equation (3) based on $r_1, r_2, S_1(t)$ and $S_2(t)$. Figure 2 shows the predicted DoR curve of the combination therapy (dashed line) when assuming the two drugs are independent, which is close to the observed DoR curve of the combination arm (blue line). The observed curve is approximately within the prediction confidence interval. The predicted median DoR for the combination arm is approximately 8 months and the observed is 6.8 months.

We evaluate the relationship of median DoRs among the combination therapy and each individual drug based on the prediction model (3). Assume a simplified case in which the two drugs are independent, i.e. $\varphi'$, $\varphi''(t)$, and $\varphi(t) = 0$. Let the medians of DoR for Drug 1 and Drug 2 be $u_1$ and $u_2$, respectively. Without losing generality, assume $u_1 < u_2$, then $S_2(u_1) > \frac{1}{2}$. Let $u$ be the median DoR of the combination therapy, then

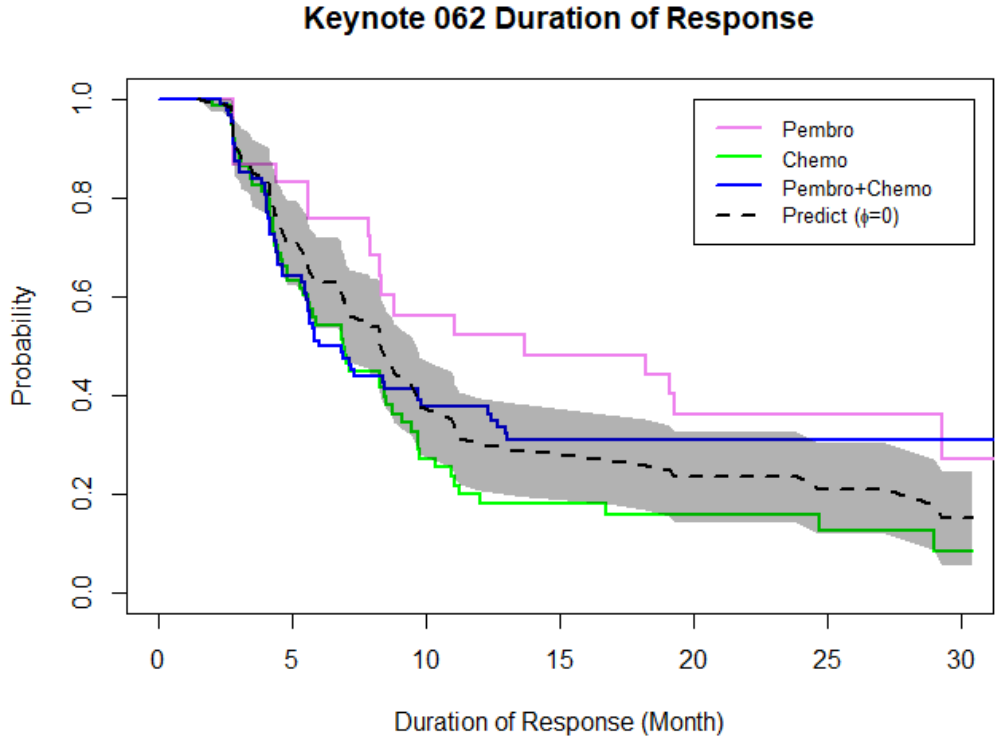

**Figure 2** Duration of response of pembrolizumab (Pembro), chemotherapy (Chemo) and the combination of pembrolizumab+chemotherapy in Keynote 062 of patients with gastric cancer. $\phi = 0$ means both $\phi'$ and $\phi''$ are zero, i.e., the prediction shown is under the assumption that the two drugs are independent.

$$S(u_1) = \frac{1}{2} \cdot \frac{r_1 + r_2(2 - r_1)S_2(u_1)}{r_1 + r_2(1 - r_1)} > \frac{1}{2} \cdot \frac{r_1 + r_2(1 - \frac{1}{2}r_1)}{r_1 + r_2(1 - r_1)} > \frac{1}{2}$$

Therefore $u > u_1$, i.e. the median DoR for the combination therapy is always longer than the shorter median DoR of the individual drug. On the other hand, it can be shown that

(i) If $S_1(u_2) < \frac{1-r_2}{2-r_2}$, then $S(u_2) < \frac{1}{2}$, i.e. $u < u_2$;

(ii) If $S_1(u_2) = \frac{1-r_2}{2-r_2}$, then $S(u_2) = \frac{1}{2}$, i.e. $u = u_2$;

(iii) If $S_1(u_2) > \frac{1-r_2}{2-r_2}$, then $S(u_2) > \frac{1}{2}$, i.e. $u > u_2$.

That is, the median DoR of the combination therapy may be longer, equal or shorter than the longer median DoR of the individual drugs. Going back to the Keynote 062 example, the probability of chemotherapy DoR longer than 13.7 months (the median DoR of the pembrolizumab arm) is about 18%, i.e. $S_1(u_2) = 18\%$, while $(1 - r_2)/(2 - r_2) = 46\%$. From above, since $S_1(u_2) < (1 - r_2)/(2 - r_2)$, the median DoR of the combination arm is expected to be shorter than the median DoR of the pembrolizumab arm. Scenario (i) in general holds true for chemotherapy and PD1/PD-L1 combination since the chemotherapy DoR is much shorter than PD1/PD-L1 DoR.

## 4. Tumor Size Change

Tumor size reduction is a direct measurement of anti-tumor effect. In oncology, the best percentage change from baseline in tumor size for target lesions is usually presented as a waterfall plot to show the tumor cell killing activity of a therapy. Note that the X-axis of a waterfall plot is usually individual patients sorted by their best tumor size changes. If the X-axis is re-scaled continuously with values from 1 to 0 and the waterfall plot is rotated 90°, the waterfall plot becomes the cumulative distribution function (CDF) curve of the best percentage change from baseline in tumor size.

Let $P_1$ and $P_2$ be a patient's best percentage tumor reduction from baseline for monotherapy Drug 1 and Drug 2, respectively, and $P$ be a patient's best percentage tumor reduction from baseline for the combination of Drug 1 and Drug 2. A positive value of $P, P_1$ and $P_2$ correspond to a reduction from baseline and a negative value correspond to an increase from baseline. To predict $P$ the tumor size reduction for the combination, we categorize best percentage tumor reduction by responses (e.g. target lesion size reduction $\geq$ 30% for solid tumors based on RECIST 1.1, or node size reduction in sum of the product of the diameters $\geq$ 50% in lymphoma based on IWG criteria [30]) into 3 types like in the DoR prediction model. Figure 3 depicts the proposed model:

- For patients who would respond to both monotherapy drugs, the combination might kill more tumor cells than the independent drug action model prediction [17]. Since tumor size reduction can be considered as proportion of tumor cell killed by the treatment, we propose to apply a modified Bliss cell-level model, which allows correlation of the drugs, for such patients. That is,

$$P = P_1 + P_2 - P_1 P_2 - \varphi\sqrt{P_1(1-P_1)P_2(1-P_2)} \qquad (4)$$

where $\varphi$ is the correlation between the two drugs. According to patient-derived tumor xenograft (PDX) studies [8, 17], it is reasonable to assume that $\varphi$ is in a range of (0, 0.3) in most combination therapies.

- For patients who would respond only to Drug 1 but not to Drug 2, we use independent drug action model, i.e. $P = \max(P_1, P_2)$. Since patients in this category only respond to Drug 1, it means $P_1 > P_2$. So $P = \max(P_1, P_2) = P_1$.

- For patients who would respond only to Drug 2 but not to Drug 1, $P = \max(P_1, P_2) = P_2$

- For patients who would respond to neither monotherapy,

$$P = \max(P_1, P_2) \qquad (5)$$

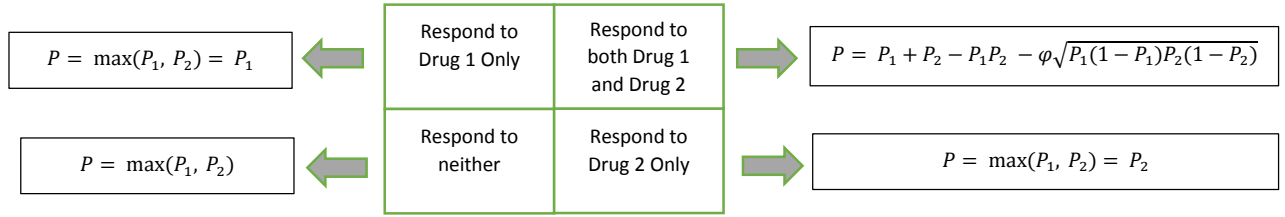

**Figure 3** Proposed model to predict the best percent tumor reduction from baseline for combination treatment

Even though the analytical form of the distribution of $P$ (i.e. CDF or equivalently waterfall plot) from the proposed model is complicated due to equation (4), sampling method can be easily implemented to predict the tumor size change of the combination therapy (R code is provided in Appendix 5). When the best % tumor size change data are available from each monotherapy drug, we correlate the two empirical distributions of $P_1$ and $P_2$ via a bivariate normal Copula. Paired samples are drawn from this joint bivariate distribution and each pair $(p_1, p_2)$ represents a patient's tumor size reduction if monotherapy Drug 1 is given and the tumor size reduction if monotherapy Drug 2 is given. If $p_1 \geq 30\%$ and $p_2 \geq 30\%$, equation (4) is applied to calculate $p$ which is the tumor size reduction for this simulated patient if combination of Drug 1 and Drug 2 is given. Otherwise, i.e. either $p_1 < 30\%$ or $p_2 < 30\%$, equation (5) is applied to calculate $p$. This way, samples of *P* and the distribution of *P* are obtained. The confidence interval of the distribution function can be attained with bootstrap method.

We apply the proposed model to Checkmate 067 [12]. Figure 4 shows the predicted waterfall plot of the combination (dashed line) assuming correlation $\varphi = 0.25$ given the waterfall plot for ipilimumab (pink line) and nivolumab (green line). The predicted curve is close to the observed combination waterfall plot (blue line), and the observed curve is generally within the confidence interval of the predicted. We assume a mild correlation of 0.25 in this example since both drugs are checkpoint inhibitors. A sensitivity analysis of prediction assuming $\varphi = 0$ is directly compared with $\varphi = 0.25$ in Appendix 6. In Palmar 2017 [17], the independent drug action model (i.e. equation (5)) was applied to all patients regardless of response types to Checkmate 067 assuming zero or 0.25 correlations. The predicted is very close to the observed waterfall plot as well, except that confidence interval of the prediction was not provided. Because the patients who would respond to both monotherapies in Checkmate 067 is approximately only 8% (product of the two monotherapy response rates 19.0% and 43.7%), the proportion of patients to which equation (4) can be applied is small. Therefore, in this case the proposed model is close to the pure independent drug action model in Palmer's paper (not shown in Figure 4).

To compare the proposed model with independent drug action model (Palmer's), we further applied the models to a Hodgkin's Lymphoma example in which large proportion of patients would respond to both monotherapies. In this example, we predict the waterfall plot of the combination therapy of nivolumab + brentuximab vedotin (BV). The monotherapy response rate of nivolumab is 65% and of BV is 75% (Response is defined as node size reduction in sum of the product of the diameters ≥ 50% from baseline in lymphoma according to IWG criteria [30]). Therefore, the proportion of patients responding to both drugs is about 50% (i.e.75%×65%=49%). It implies that equation (4) can be applied to about half of the patients in the proposed model. In Figure 5, the waterfall plots for nivolumab monotherapy (green line) (n=95) [27], BV monotherapy (pink line) (n=98) [28], and combination of the two from a recent Phase 1/2 study (blue line) (n=17) [29] are obtained by digitizing three separate waterfall plots from the original publications. The predicted waterfall plot for the combination (assuming $\varphi = 0$) from the proposed model is shown as the black dashed line and from the independent action model is shown as the grey dashed line ("Palmer's"). The prediction for node size reduction < 50% is the same for the proposed model and Palmer's model, and the prediction for node size reduction > 50% from the proposed model is closer to the observed waterfall plot than the prediction from Palmer's model. While we assume a zero-correlation due to different mechanism of action for nivolumab and BV, a sensitivity analysis under $\phi = 0.25$ is compared between the two methods (Appendix 6 Figure S2). Both correlation assumptions indicate the proposed method outperforms Palmer's for this case. The limitation of this example is the small sample size of the observed combination data. After more trial data are published, further investigation of the proposed model when large proportion of patients respond to both monotherapies can be conducted.

Prediction of tumor size change for the combination is important because it can shed light on the depth of the response to combination therapy compared to monotherapies in addition to the increased response rate in the traditional definition using RECIST 1.1. Importantly, not only the response rate but also the depth of response may correlate with progression-free survival (PFS) and overall survival (OS) [9, 15]. As well, with recent advances of new treatments, the response rates from standard of care (SOC) in many first line cancer types (e.g. renal cell carcinoma, small cell lung cancer, bladder cancer, etc.) are very high. Some are even in the 60% - 70% range which implies that there may not be much room left for improvement in terms of response rate. This calls for emerging endpoints, e.g. deep response rate, for Phase II proof-of-concept (PoC) studies. The conventional complete response (CR) per RECIST 1.1 can be considered a deep response as the most extreme scenario in which tumor reduction of 100% is used as the criterion. Similarly, one can use another cutoff, e.g. tumor reduction ≥ 75%, as the criterion to define deep response.

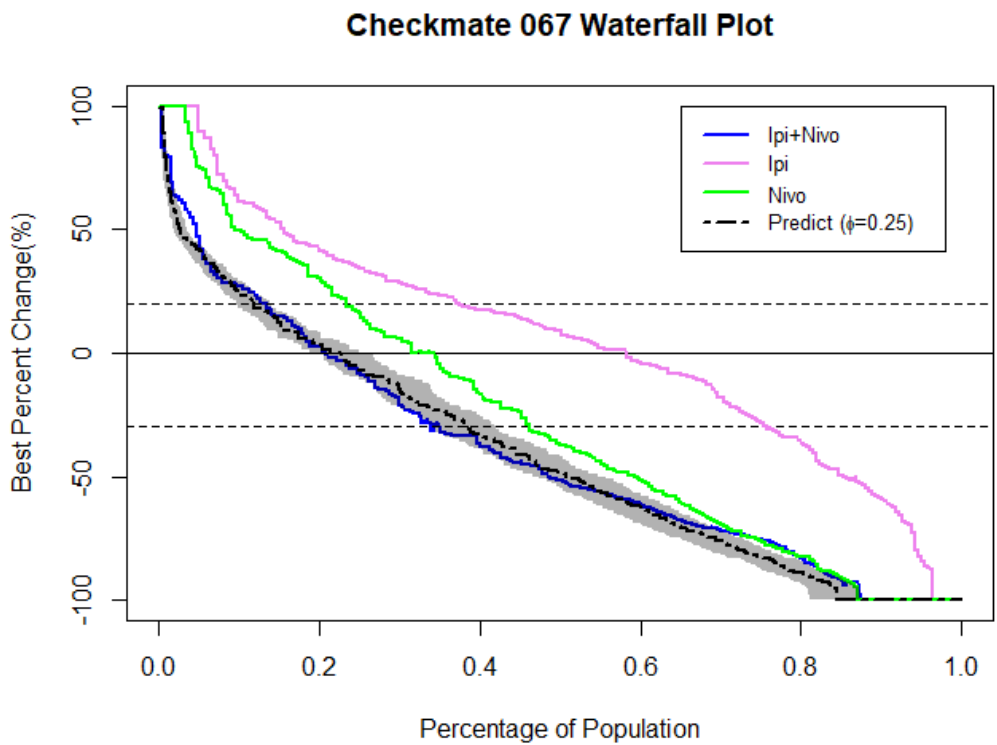

**Figure 4** Best % tumor size change (a positive value means tumor size increase, and a negative value means tumor reduction) of ipilimumab (Ipi), nivolumab (Nivo) and the combination of ipilumumab plus nivolumab in Checkmate 067 of melanoma patients, assuming mild correlation 0.25.

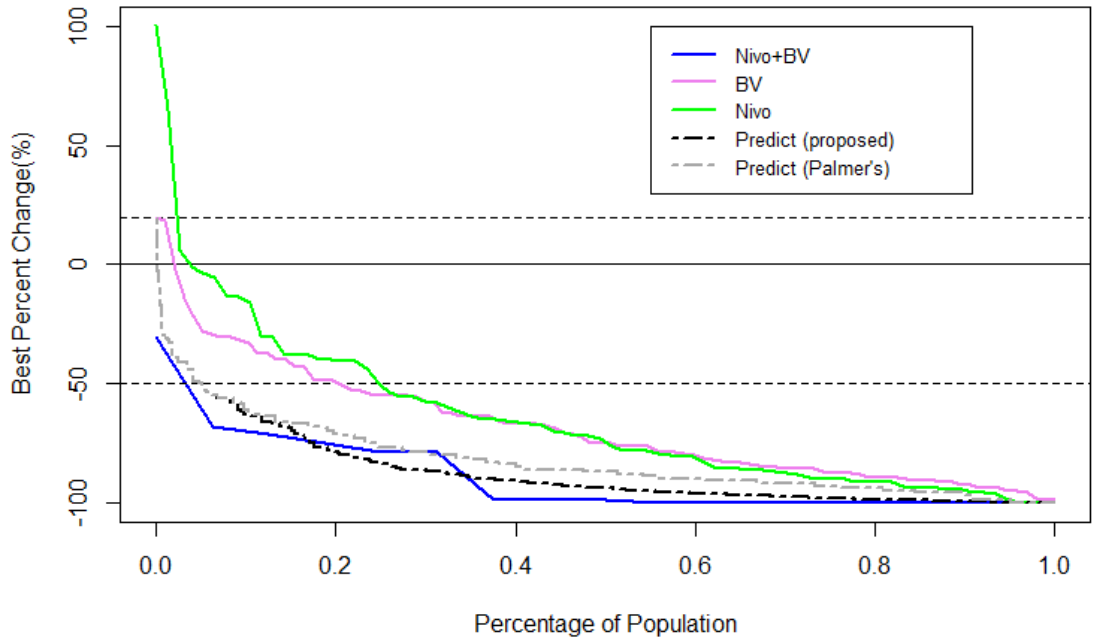

**Figure 5** Best % node size change in sum of the product of diameters (a positive value means size increase, and a negative value means reduction) of nivolumab (Nivo), brentuximab vedotin (BV) and the combination of nivolumab plus BV in R/R Hodgkin Lymphoma patients, assuming zero correlation.

In a hypothetical example, monotherapy data of Drug 1 and Drug 2 are available. Drug 1 is the standard of care (SOC) with response rate as high as 70%, and the monotherapy data is from a large Phase 3 trial. Drug 2 is a new experimental therapy with response rate of 35%, and the monotherapy data is from a single arm Phase I/II trial. According to equation (1) the combination response rate is predicted to be about 80%, i.e. only 10% increase from Drug 1. If we were to design a randomized Proof-of-Concept (PoC) study to compare the combination with the SoC using response rate as the primary endpoint, to have 80% power more than 400 patients are needed with 5% alpha (one-sided), which is not a practical size for a Phase II PoC study. Alternatively, we consider deep response rate as the primary endpoint for this PoC study. For the sample size and power calculation, we first need to evaluate the expected increase in deep response rate which the combination can improve over SoC. We apply the proposed model to predict the waterfall plot for the combination based on the two monotherapy waterfall plots as shown in Figure 6. The predicted waterfall plot shows that the combination can not only increase response rate but can also induce deeper tumor reduction, which implies that the improvement in deep response rate may be more evident than in response rate. In fact, if deep response is defined as tumor size reduction ≥ 75%, then the deep response rate of Drug 1 is 40% and the predicted deep response rate of the combination is 60% from the waterfall plots. To have 80% power to detect deep response rate difference of 60% vs. 40%, approximately 200 patients are needed with 5% alpha (one-sided), which is a more reasonable size for a Phase II PoC study. Through this hypothetical example, we illustrate the use of prediction of tumor size change of the combination therapy to guide the study design and endpoint selection of oncology drug development.

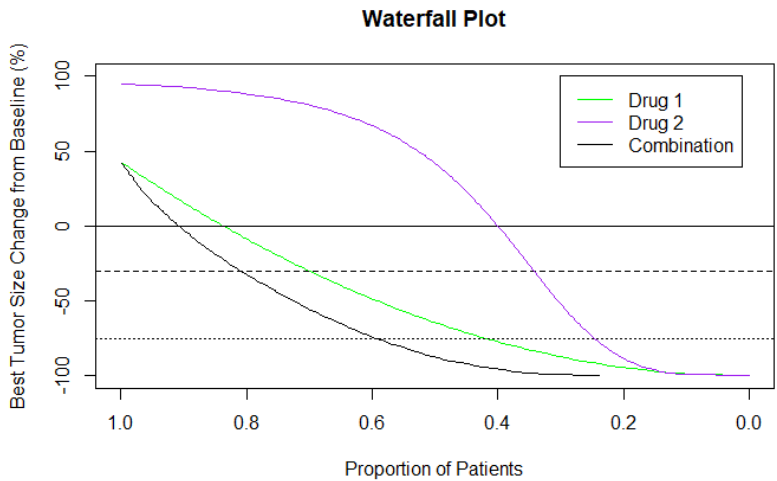

**Figure 6** Waterfall plot of individual Drug 1 and Drug 2 and predicted waterfall plot for the combination.

## 5. Summary and Discussion

In this paper we propose simple models to predict duration of response and tumor size reduction for combination therapies, based on the independent action concepts. The model for predicting DoR is a new application of the independent drug action concept, and the model for predicting tumor size change refines the prediction for patients who would respond to both individual drugs from the independent drug action model. Additionally, presentation of confidence intervals is a significant improvement from Palmer 2017 [17] and Palmer 2020 [18].

The two examples (Keynote 062 and Checkmate 067) presented here use monotherapy data from the same trial to predict the combination effect. Particularly for Checkmate 067 the ipilimumab plus nivolumab combination has now been shown to fit the independent activity model for PFS (Palmer), ORR (Schmidt) and best tumor size change (Palmer and Figure 4). In previous publications, monotherapy data were usually not from the same trial with randomization, but from two separate trials. It is likely that the patient characteristics are different between any two trials, so confirmation of the independent drug action model within one randomized trial is important. If the monotherapy data are from two trials, propensity score matching or weighting methods [1, 13] for adjusting populations could be applied to estimate the monotherapy efficacy before the prediction of combination efficacy. Alternatively, machine learning techniques could be used to estimate individual patient's counter factual treatment effect (ITE) to adjust the confounding covariates [14] before prediction of the combination efficacy. In addition, the follow up time in two separate trials of the monotherapies may be very different. In such scenarios, when predicting the waterfall plot for the combination treatment, we need to adjust the best tumor-size-change to equivalent follow-up times from any two monotherapy trials.

With the approval of PD-1 checkpoint inhibitors, the treatment paradigm changes substantially for many tumor indications. New combinations are being developed to address the high unmet medical need for patients who progressed from PD-1 checkpoint inhibitors. If the new therapy is the combination of a new drug and a PD-1 checkpoint inhibitor, the monotherapy of the new drug and the combination can be evaluated in PD-1/PD-L1 exposed population, but it may not be ethical to test the monotherapy PD-1 checkpoint inhibitor again in this population. Then what is the potential effect of the monotherapy PD-1 checkpoint inhibitor in this combination for PD-1/PD-L1 exposed population? For PFS endpoint, Chen 2020 [4] gives the formula to reverse-engineer and estimate the monotherapy effect from the combination and the other monotherapy. Similarly, the reverse-engineering can be done for other endpoints. For example, if the response rate of the combination ($r$) and the monotherapy new drug ($r_1$) are known, then the possible response rate of monotherapy PD-1 checkpoint inhibitor ($r_2$) can be estimated from equation (1). Such estimation for response duration and tumor size change may not be as straightforward as with response rate though. Smoothing or monotonic restrictions may be needed. It is a topic of ongoing investigation.

Although, "All models are wrong, but some are useful" (George E. P. Box), the models introduced in this paper are meant to serve as base-case models, in the absence of alternatives, to facilitate strategic planning and decision making in early stage oncology drug development..

# Appendix 1

Derivation of equation (1)

$$\varphi' = \frac{Cov(X_1, X_2)}{\sqrt{var(X_1)var(X_2)}} = \frac{E(X_1 X_2) - E(X_1)E(X_2)}{\sqrt{var(X_1)var(X_2)}} = \frac{\Pr(X_1 = 1 \text{ and } X_2 = 1) - r_1 r_2}{\sqrt{r_1(1-r_1)r_2(1-r_2)}}$$

Therefore,

$$\Pr(X_1 = 1 \text{ and } X_2 = 1) = r_1 r_2 + \varphi'\sqrt{r_1(1-r_1)r_2(1-r_2)}$$

Then

$$\Pr(X_1 = 1 \text{ or } X_2 = 1) = \Pr(X_1 = 1) + \Pr(X_2 = 1) - \Pr(X_1 = 1 \text{ and } X_2 = 1)$$

$$= r_1 + r_2 - r_1 r_2 - \varphi'\sqrt{r_1(1-r_1)r_2(1-r_2)}$$

# Appendix 2

The relationship of $\varphi(t)$, $\varphi'$, and $\varphi''(t)$ in equation (4). In the following derivation, we drop the $(t)$ in those functions of $t$ for simplification.

$\varphi'$ is the correlation between $X_1$ and $X_2$, that is $\varphi' = corr(X_1, X_2)$.

$\varphi''(t)$ is the correlation between indicator $I_{T_1>t}$ and $I_{T_2>t}$. For simplification, let $I_1 = I_{T_1>t}$ and $I_2 = I_{T_2>t}$. Then $\varphi'' = corr(I_1, I_2)$.

$\varphi(t)$ is the correlation between indicator $I_{X_1 T_1>t}$ and $I_{X_2 T_2>t}$. Notice that $I_{X_1 T_1>t} = X_1 I_1$ and $I_{X_2 T_2>t} = X_2 I_2$.

Therefore,

$$\varphi = corr(X_1 I_1, X_2 I_2) = \frac{E(X_1 I_1 X_2 I_2) - E(X_1 I_1)E(X_2 I_2)}{\sqrt{var(X_1 I_1)var(X_2 I_2)}} = \frac{E(X_1 X_2)E(I_1 I_2) - r_1 S_1 r_2 S_2}{\sqrt{r_1 S_1(1-r_1 S_1)r_2 S_2(1-r_2 S_2)}}$$

$$= \frac{(r_1 r_2 + \varphi'\sqrt{r_1(1-r_1)r_2(1-r_2)})(S_1 S_2 + \varphi''\sqrt{S_1(1-S_1)S_2(1-S_2)}) - r_1 S_1 r_2 S_2}{\sqrt{r_1 S_1(1-r_1 S_1)r_2 S_2(1-r_2 S_2)}}$$

# Appendix 3

The variance of the predicted survival function of DoR of the combination therapy:

$$Var(\hat{S}(t)|r_1, r_2, \varphi, \varphi') \approx \frac{r_1^2\left((1-r_2S_2) - \varphi\frac{A_2}{2}B^{-\frac{1}{2}}\right)^2 \sigma_{S_1}^2 + r_2^2\left((1-r_1S_1) - \varphi\frac{A_1}{2}B^{-\frac{1}{2}}\right)^2 \sigma_{S_2}^2}{r^2},$$

where

$A_1 = r_2 S_2 (1 - r_2 S_2)(1 - 2r_1 S_1),$

$A_2 = r_1 S_1 (1 - r_1 S_1)(1 - 2r_2 S_2),$

and $B = r_1 r_2 S_1 S_2 (1 - r_1 S_1)(1 - r_2 S_2).$

# Appendix 4

**Table 1** Duration of Response of Combinations of Chemotherapy and anti-PD1/PD-L1

| Population | Drug1 | Drug1 Median DoR (month) | Drug2 | Drug2 Median DoR (month) | Combination Median DoR (month) |
|---|---|---|---|---|---|
| 1L NSCLC non-squamous | Chemotherapy | 7.8 [7] | Pembrolizumab | 16.8 [6] | 11.2 [7] |
| | Bevacizumab + chemotherapy | 5.7 [24] | Atezolizumab | 16.3 [20] | 9.0 [24] |
| | chemotherapy | 6.1 [25] | Atezolizumab | 16.3 [20] | 8.4 [26] |
| 1L NSCLC squamous | Chemotherapy | 4.8 [19] | Pembrolizumab | 16.8 [5] | 7.7 [19] |
| 1L TNBC | Chemotherapy | 5.6 [21] | Atezolizumab | 21 [5] | 7.4 [21] |

| 1L Gastric Cancer CPS1 | Chemotherapy | 6.8 [23] | Pembrolizumab | 13.7 [23] | 6.8 [23] |

# Appendix 5

1. R code of copula method to predict best percent tumor size change in combination therapy.

```
library(MASS)
copula.pred <- function(m=2, n=5000, rho=0.25, input1=pembro$PCHG,
input2=chemo$PCHG, cutoff=-30, x.grid=seq(-120, 100, by=1), palmer=FALSE){

# n is the number of samples drawn

  # the empirical CDF of the monotherapy best % tumor size change. ewcdf(.) can be used
if we want to weigh the subjects with propensity scores #

 Fn <- ecdf(input1)
 y1 <- Fn(x.grid)

 Fn2 <- ecdf(input2)
 y2 <- Fn2(x.grid)

# draw random samples of the correlated CDFs from Gaussian copula

 sigma <- matrix(c(1, rho, rho, 1),nrow=2)
 z <- mvrnorm(n, mu=rep(0, m),Sigma=sigma, empirical=T)

 u <- pnorm(z)

 samp1 <- x.grid[findInterval(u[,1],y1)]
 samp2 <- x.grid[findInterval(u[,2],y2)]

# readjust to exclude samples with percent change beyond -100

 samp11 <- ifelse (samp1< -100, -100, samp1)
```

```r
       samp22 <- ifelse (samp2< -100, -100, samp2)

   # Apply the proposed model to predict best percent tumor size change for the
combination

   samp <- data.frame(out1=samp11, out2=samp22)
   min <- apply(samp, 1, min)
   if(palmer==TRUE){ # Palmer's approach
     samp$pred <- min
   } else {
     p1 <- abs(samp[,1]/100) # absolute percent reduction
     p2 <- abs(samp[,2]/100)
     sum <-  100*(p1+p2-p1*p2-rho*sqrt(p1*(1-p1)*p2*(1-p2)))
     rs <- as.numeric(samp$out1< cutoff & samp$out2< cutoff) # apply the cutoff value to
define responders, -30 for unidimensional measures and -50 for sum of product of the
axis#

     samp$pred <- ifelse(rs==0, min, -sum)
   }

   samp<- samp[order(samp$pred, decreasing = T),]
   samp$index <-  seq(0,1,length=nrow(samp))
   return(samp)
 }
```

2. R code to create bootstrap confidence interval for predicted tumor size change.

```r
   boot.out <- function(nboot=2000, nsim=5000, num=NULL, input1, input2, cutoff=-30, rho,
palmer=FALSE){
     boot.pred <- matrix(NA, nrow=nsim, ncol=nboot)
     for (i in 1:nboot){
       in1 <- sample(input1, size=ifelse(is.null(num),length(input1),num), replace=T)
       in2 <- sample(input2, size=ifelse(is.null(num),length(input2),num), replace=T)
       boot.pred[, i] <- copula.pred(m=2, n=nsim, rho=rho, input1=in1, input2=in2,
cutoff=cutoff, x.grid=seq(-120, 100, by=1), palmer=palmer)$pred
     }
     lower.ci <- apply(boot.pred, 1, quantile, probs=0.05)
     upper.ci <- apply(boot.pred, 1, quantile, probs=0.95)
     mean.boot <- apply(boot.pred, 1, mean)
     out <- cbind(mean=mean.boot, lower.ci=lower.ci, upper.ci=upper.ci)
     out <- as.data.frame(out)
     out$index <-  seq(0,1,length=nrow(out))
     return(out)
   }
```

**Appendix 6**

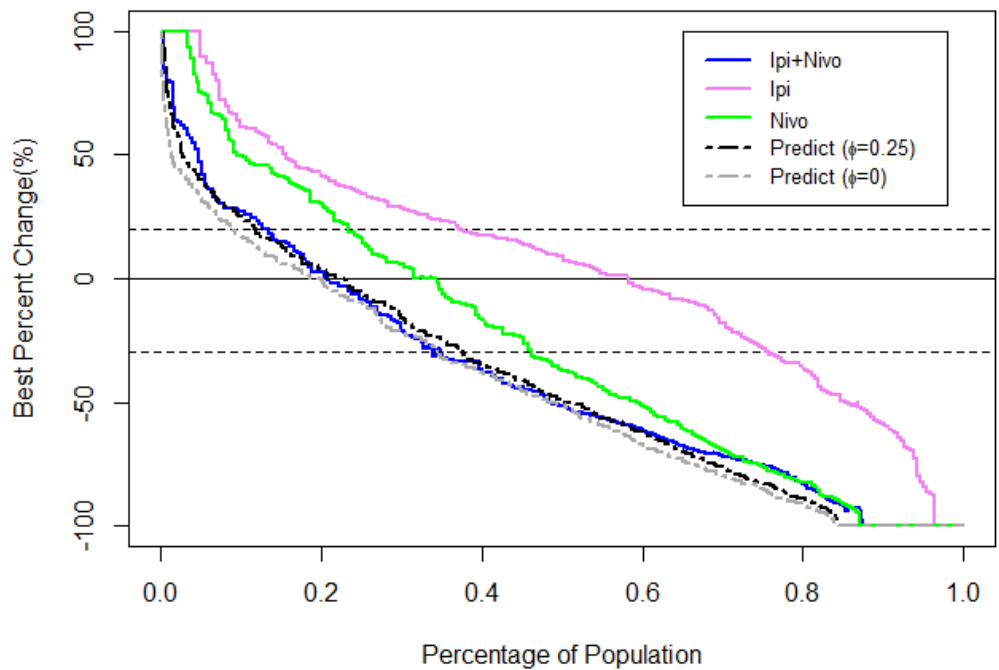

**Figure S1** Best % tumor size change of ipilimumab, nivolumab and the combination of ipilumumab+nivolumab in Checkmate 067 of melanoma patients under correlation = 0.25 and correlation = 0.

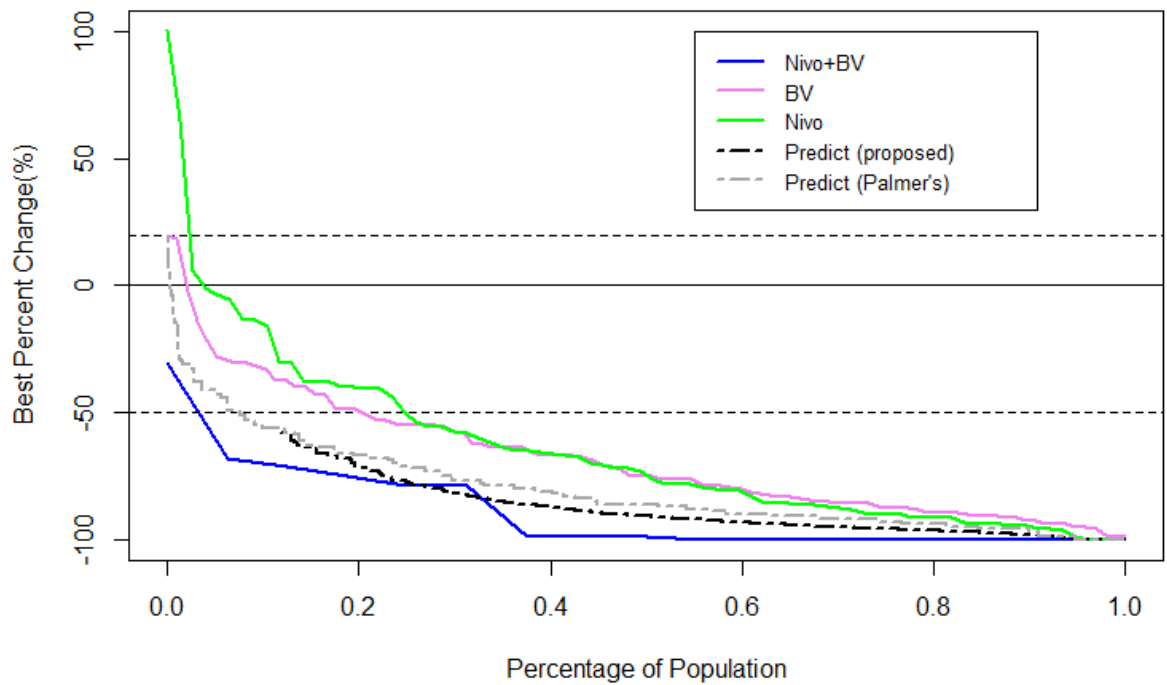

**Figure S2** Best % node size change in sum of the product of diameters (a positive value means size increase, and a negative value means reduction) of nivolumab (Nivo), brentuximab vedotin (BV) and the combination of nivolumab plus BV in R/R Hodgkin Lymphoma patients, under correlation of 0.25 for proposed approach and Palmer's.